\documentclass[twocolumn,10pt,amsmath,amssymb,superscriptaddress,prb,floatfix,aps]{revtex4-1}
\usepackage{graphicx}
\usepackage{bm}

\usepackage[colorlinks=true, citecolor=red, linkcolor=blue, urlcolor=black]{hyperref}
\usepackage[english]{babel}
\newcommand{\ket}[1]{\left|#1\right\rangle}
\newcommand{\bra}[1]{\left\langle#1\right|}

\begin{document}

\title{Quantum walk versus classical wave:  Distinguishing ground states of quantum magnets by spacetime dynamics}

\author{Piotr Wrzosek}
\affiliation
{Faculty of Physics, University of Warsaw, Pasteura 5, PL-02093 Warsaw, Poland}

\author{Krzysztof Wohlfeld}
\email{krzysztof.wohlfeld@fuw.edu.pl}
\affiliation
{Faculty of Physics, University of Warsaw, Pasteura 5, PL-02093 Warsaw, Poland}

\author{Damian Hofmann}
\affiliation 
{\mbox{Max Planck Institute for the Structure and Dynamics of Matter, Luruper Chaussee 149, D-22761 Hamburg, Germany}}

\author{Tomasz Sowi\'nski}
\email{tomasz.sowinski@ifpan.edu.pl}
\affiliation
{\mbox{Institute of Physics, Polish Academy of Sciences, Aleja Lotnik\'ow 32/46, PL-02668 Warsaw, Poland}}

\author{Michael A.~Sentef}
\email{michael.sentef@mpsd.mpg.de}
\affiliation 
{\mbox{Max Planck Institute for the Structure and Dynamics of Matter, Luruper Chaussee 149, D-22761 Hamburg, Germany}}

\date{\today}

\begin{abstract}
We investigate the wavepacket spreading after a single spin flip in prototypical two-dimensional ferromagnetic and antiferromagnetic quantum spin systems. We find characteristic spatial magnon density profiles: While the ferromagnet shows a square-shaped pattern reflecting the underlying lattice structure, as exhibited by quantum walkers, the antiferromagnet shows a circular-shaped pattern which hides the lattice structure and instead resembles a classical wave pattern. We trace these fundamentally different behaviors back to the distinctly different magnon energy-momentum dispersion relations and also provide a real-space interpretation. Our findings point to new opportunities for real-time, real-space imaging of quantum magnets both in materials science and in quantum simulators. 
\end{abstract}
\maketitle

\section{Introduction}
Two-dimensional quantum magnets are quintessential quantum many-body systems that come in two main realizations: antiferromagnets (AF) or ferromagnets (FM). AF are prototypical condensates (BCS superconductors, superfluids, crystals), in which classical order is dressed by its associated quantum fluctuations
 \cite{khomskii_basic_2010}.
Whereas the latter do not destroy the order at $T=0$ -- as would happen for AF chains, in agreement with the Coleman theorem -- the quantum reduction of the order parameter is of the order of 40\%. Such strong quantum effects are intrinsically related to the onset of the low-lying excitations above the respective ground state (Goldstone modes), which are coined magnons and have linear-in-momentum ($|\bm{k}|$) quasi-particle dispersion.
By contrast, FM can be regarded as more unique because their fully polarized ground state does not contain any quantum fluctuations, and the low-lying excitations disperse as $\bm{k}^2$. 
Hence the FM ground state can be viewed a natural realization of a true vacuum, and the associated magnon excitations as particles.

\begin{figure}
	\centering
	\includegraphics[width=\columnwidth]{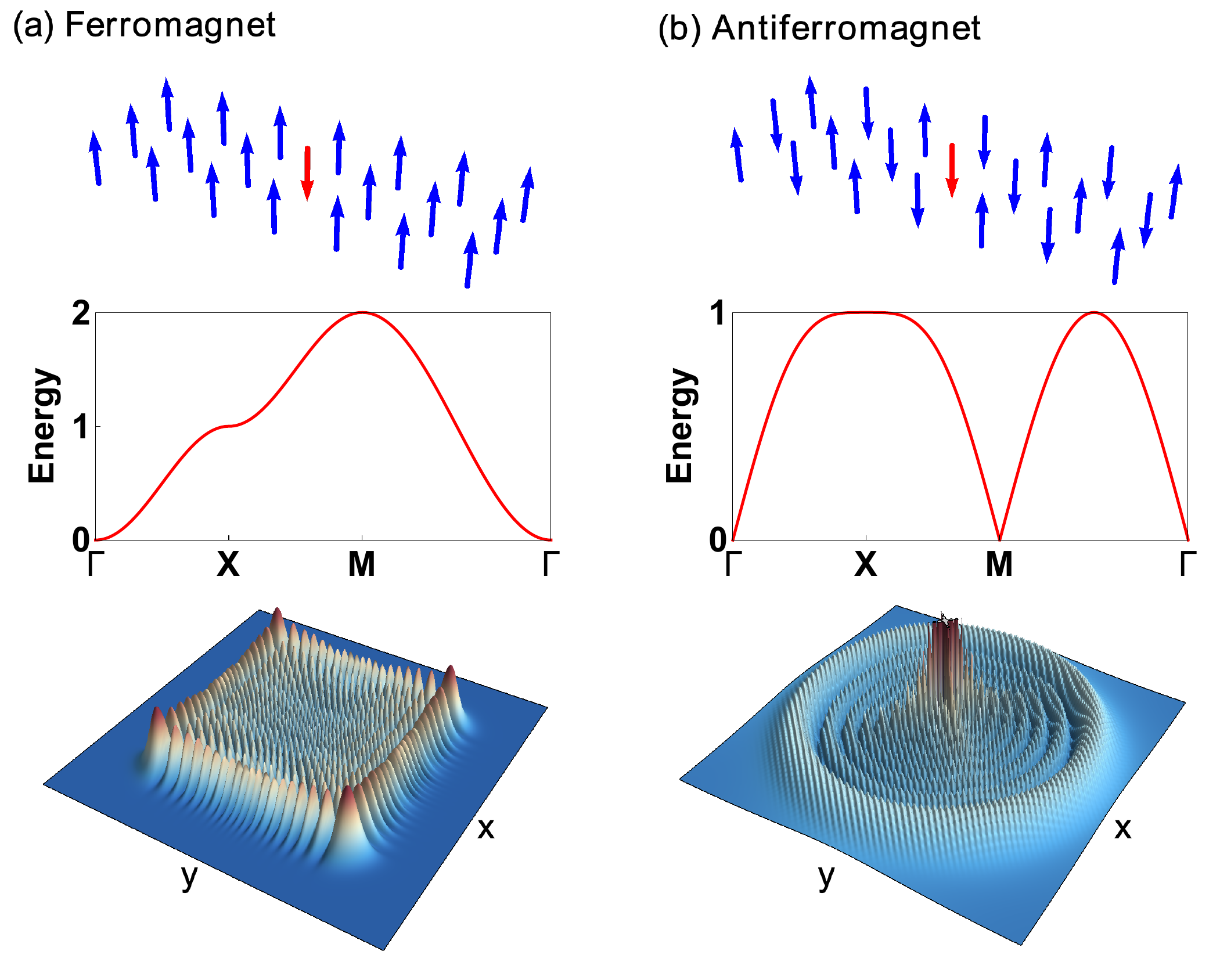}    
	\caption{{\bf Single-magnon excitation properties in two-dimensional quantum magnets.} (a) Ferromagnet, (b) antiferromagnet. Top row: Cartoon of the ground state with a single spin flip (red). Middle row: Dispersion relation of single-magnon excitations (in energy units where $J=1$), with $\bm{k^2}$ (FM) and $|\bm{k}|$ (AF) low-energy behavior around $\Gamma$, respectively. Bottom row: Snapshot of spatial density profile for $t \sim 200$ fs ($30 \hbar/J$) after spin-flip excitation. 
    To arrive at this time scale, spin exchange is taken to be a representative value of $J=100$ meV.
	}\label{fig:1}
\end{figure}

In traditional condensed matter physics the questions of magnetic ground states and their associated low-lying excitations on the atomic length scale are investigated experimentally with scattering techniques (neutrons, X-rays), which yield information in reciprocal space (momentum $\bm{k}$, frequency $\omega$). On the other hand, tremendous progress in controlling ultracold gases in optical lattices has provided a complementary real-space and real-time ($\bm{r},t$) perspective on archetypal spin Hamiltonians~\cite{2008BlochRMP,LewensteinBook}. Due to the tunability of these systems, it is now possible to perform quantum simulations of systems described by celebrated Hamiltonians previously considered as minimal toy models, such as the fermionic \cite{2008SchneiderScience,2008JordensNature} and bosonic \cite{2002GreinerNature} Hubbard model, the Ising model \cite{2011SimonNature}, and the Heisenberg model \cite{2013YanNature}. The spacetime-resolved microscopic imaging of such quantum simulators is possible thanks to the single-site fluorescence imaging technique invented almost decade ago \cite{2009BakrNature,2010ShersonNature,2010BakrScience}, and further developed recently \cite{2011WeitenbergNature,2015HallerNatPhys,2015EdgePRA,2016ParsonsScience,2016BollScience,2016CheukScience}. In particular, this technique was successfully used for the quantum simulation and the spacetime probing of AF order in a two-dimensional lattice \cite{2017MazurenkoNature}.

In this work, we take a fresh look at the old problem of magnetic ground states and their low-lying excitations focussing on generic 2D square-lattice quantum magnets. We examine the spacetime dynamics of a single initially localized excitation on top of the respective magnetic ground state. We find simple yet remarkable and robust distinguishing fingerprints between the FM and AF cases. In the FM case the problem is readily mapped onto the problem of a single quantum particle in the vacuum. Thus it is classified as the well-known quantum walk in continuous time on a discrete spatial lattice, which recently is under extensive theoretical and experimental exploration~\cite{2009KarskiScience,2009SchmitzPRL,2010PeruzzoScience,2010ZahringerPRL,Venegas-Andraca2012,2012LahiniPRA,2015PreissScience,2017WiaterPRA,2019YePRL,2019YanScience,2020MondalARX}. As expected from intuition based on this analogy, a square pattern emerges in the spatial density profile after excitation, reflecting the underlying crystal symmetry [Fig.~\ref{fig:1}(a)]. By striking contrast, the dynamics above the AF ground state is instead reminiscent of classical wave, with isotropic circular patterns largely ignorant of the crystal symmetry [Fig.~\ref{fig:1}(b)]. We trace this quantum-walk versus classical-wave behavior back to the fundamental difference in quantum ground states and their associated low-energy excitations. 

\section{Model: ground and excited states}
%
Consider the 2D spin $S=\frac12$ Heisenberg model
\begin{equation}
    \hat{\cal H} = J \sum_{\langle i,j\rangle} \hat{\bm{S}}_{\bm{r}_i} \cdot \hat{\bm{S}}_{\bm{r}_j}
\end{equation}
for spins $S=\frac12$ on a 2D square lattice with nearest-neighbor spin exchange coupling $J>0$ ($J<0$) in case of antiferromagnet (ferromagnet) respectively. 

In order to introduce magnons via the standard Holstein-Primakoff transformation in the AF case we first rotate all spins on sublattice $B$ of the AF state: We assume that the corresponding Neel state of the AF ground state in question is such that all spins are up (down) on sublattice $A$ ($B$), respectively. (In the FM we keep the spins intact). Next we define the following Holstein-Primakoff transformation, which already contains the linear spin-wave theory (LSW) approximation \cite{Auerbach1994,Mou10}:
\begin{subequations}
\begin{align}
\hat{S}^z_{\bm{r}_j} &= \frac12 - \hat{a}_{\bm{r}_j}^{\dagger} \hat{a}_{\bm{r}_j},  \\
\hat{S}^+_{\bm{r}_j} &\approx \hat{a}_{\bm{r}_j},  \\
\hat{S}^-_{\bm{r}_j} &\approx \hat{a}_{\bm{r}_j}^{\dagger},
\end{align}
\end{subequations}
followed by Fourier and Bogolyubov transformations, with the latter defined as
\begin{equation}\label{eq:bog_tr}
\hat{a}_{\bm{k}} = u_{\bm{k}}\hat{\alpha}_{\bm{k}} + v_{\bm{k}}\hat{\alpha}_{\bm{-k}}^{\dag}.
\end{equation}
we obtain a diagonal form in terms of the Bogolyubov magnons,
\begin{equation}\label{eq:hfinal}
\hat{H} = \sum_{\bm{k}} \omega_{\bm{k}} \hat{\alpha}_{\bm{k}}^{\dag} \hat{\alpha}_{\bm{k}} +\frac{1}{2}\sum_{\bm{k}} \left(\omega_{\bm{k}} - A_{\bm{k}} \right),
\end{equation}
where
\begin{equation}\label{eq:bog}
A_{\bm{k}} = 2J\left(1 -\left(1 -\frac{J}{|J|}\right)\frac{\gamma_{\bm{k}}}{2}\right), \quad u_{\bm{k}}^2, v_{\bm{k}}^2 = \frac{A_{\bm{k}} \pm \omega_{\bm{k}}}{2\omega_{\bm{k}}},
\end{equation}
with the energy of the Bogolyubov magnons given as
\begin{equation}
\omega_{\bm{k}} = 2J\sqrt{\left(1-\gamma_{\bm{k}}\right)\left(1+\frac{J}{|J|}\gamma_{\bm{k}}\right)},
\end{equation}
and $\gamma_{\bm{k}} = \frac{1}{2}(\cos k_x + \cos k_y)$.
\section{Magnon density profiles: definition and equations}
The main goal of the paper is to investigate how a single spin flip excitation on a given site $\bm{r}_0$ on top of the ground state propagates in space and time. To this end, we calculate here the space-time dependence of a density profile $\rho\left(\bm{r},t\right)$ of a single spin-flip excitation in a quantum magnet:
\begin{subequations}
\begin{align}\label{eq:rho_spin}
\rho\left(\bm{r},t\right) &= -\bra{\varnothing} \hat{S}^+_{\bm{r}_0} \hat{S}^z_{\bm{r}_i}\left(t\right)  \hat{S}_{\bm{r}_0}^{-}\ket{\varnothing},
\\
 \hat{S}^z_{\bm{r}_i} \left(t\right) & = e^{i\frac{\hat{H}}{\hbar}t}  \hat{S}^z_{\bm{r}_i}  e^{-i\frac{\hat{H}}{\hbar}t},
\end{align}
\end{subequations}
with
$\bm{r} = \bm{r}_i - \bm{r}_0$.
This equation defines the following protocol for the dynamics.
Starting from the ground state $\ket{\varnothing}$ as the initial state of the Hamiltonian $\hat{\cal H}$, we apply at time $t=0$ a single spin-flip operator locally on site at $\bm{r}_0\equiv 0$, which is assumed to belong to the $A$ sublattice (see above; note that, that this choice does not restrict the validity of the results below but simplifies the notation). Next, at an arbitrary later time time $t$ we measure the magnetisation at site ${\bm{r}_i}$ and obtain the spatiotemporal profile of the single spin flip excitation.

We now rewrite the above protocol in terms of the Holstein-Primakoff magnons that are subject to the LSW approximation (i.e. noninteracting). In this case, we start from the ground state of the LSW-approximate Hamiltonian \eqref{eq:hfinal}, which is given by the magnon vacuum $\ket{\varnothing_{\alpha}}$. We note that this is the exact ground state (fully polarized state) for the FM case, while it is the approximate ground state for the AF case as it neglects additional quantum fluctuations caused by magnon-magnon interactions. 
Next, as in the LSW approximation a single spin-flip amounts to creating a single magnon and the magnetisation to the magnon density we obtain the following spatiotemporal excitation density profile written in terms of the noninteracting bosons:
\begin{subequations}
\begin{align}
\rho\left(\bm{r},t\right) &= \bra{\varnothing_{\alpha}} \hat{a}_{\bm{r}_0} \hat{n}_{\bm{r}_i} \left(t\right) \hat{a}_{\bm{r}_0}^{\dag} \ket{\varnothing_{\alpha}},	\\
 \hat{n}_{\bm{r}_i} \left(t\right) & = e^{i\frac{\hat{H}}{\hbar}t}  \hat{n}_{\bm{r}_i}  e^{-i\frac{\hat{H}}{\hbar}t},
\end{align}
\end{subequations}
where we skipped the constant terms.
The above equation can be understood as the time-dependent expectation value of the magnon density operator of a state with a single magnon created at a particular site at the initial time $t=0$.

\begin{figure*}[t!]
	\centering
	\includegraphics[width=\textwidth]{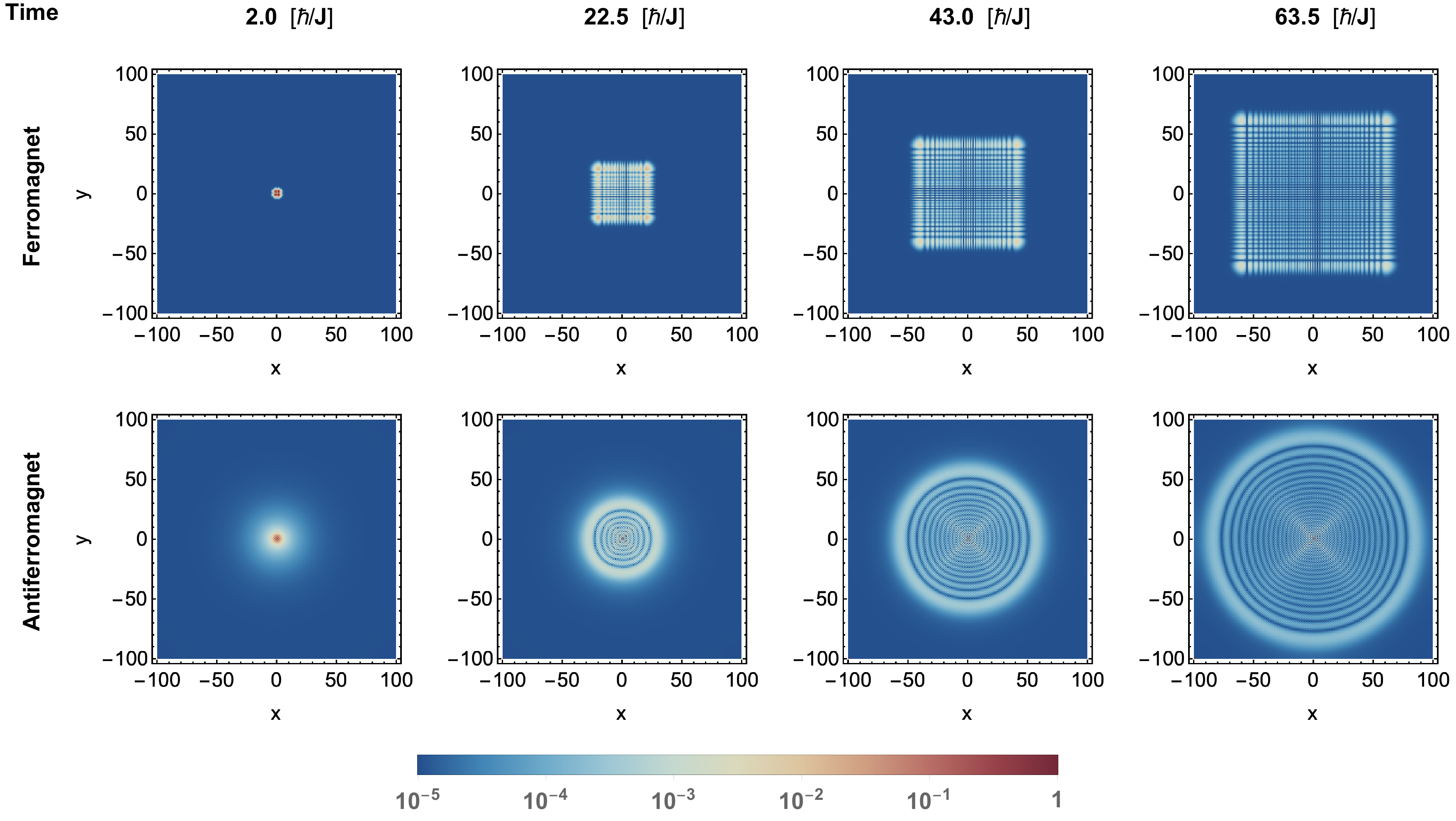}    
	\caption{{\bf Snapshots of spatiotemporal evolution of the density profiles.} Panels show the respective real-space density profiles after spin flip at position $(x=0, y=0)$ and time $t=0$ for the ferromagnet (top row) and antiferromagnet (bottom row), respectively. Columns correspond to different waiting times after excitation as indicated.
	}\label{fig:2}
\end{figure*}

We now perform a few manipulations, in order to obtain an explicit expression for 
$\rho\left(\bm{r},t\right)$. First, we rewrite:
\begin{equation}\label{eq:rho_def}
\rho\left(\bm{r},t\right) =  \langle\phi \left(\bm{r},t\right)  \ket{\phi \left(\bm{r},t\right) },
\end{equation}
where
\begin{equation}
\ket{\phi \left(\bm{r},t\right) } = \hat{a}_{\bm{r}_i} e^{-i\frac{\hat{H}}{\hbar}t} \hat{a}_{\bm{r}_0}^{\dag} \ket{\varnothing_{\alpha}}.
\end{equation}
Applying Fourier transform, $ \hat{a}_{\bm{r}} = \frac{1}{\sqrt{N}}\sum_{\bm{k}}e^{-i\bm{k}\bm{r}}\hat{a}_{\bm{k}}$ to $\ket{\phi \left(\bm{r},t\right) }$ we have,
\begin{align}
\ket{\phi \left(\bm{r},t\right) } =
\frac{1}{N}\sum_{\bm{k},\bm{k'}}e^{-i(\bm{k}\bm{r}_i - \bm{k'}\bm{r}_0)} \hat{a}_{\bm{k}} e^{-i\frac{\hat{H}}{\hbar}t} \hat{a}_{\bm{k'}}^{\dag} \ket{\varnothing_{\alpha}}.	
\end{align}
After Bogolyubov transform (\ref{eq:bog_tr}) the above appears in the form,
\begin{align}\label{eq:rh_dens}
\ket{\phi \left(\bm{r},t\right) } &=
\frac{1}{N}\sum_{\bm{k},\bm{k'}}e^{-i(\bm{k}\bm{r}_i - \bm{k'}\bm{r}_0)} (u_{\bm{k}}\hat{\alpha}_{\bm{k}} + v_{\bm{k}}\hat{\alpha}_{\bm{-k}}^{\dag})~\times \nonumber
\\
&\times e^{-i\frac{\hat{H}}{\hbar}t} (u_{\bm{k'}}\hat{\alpha}_{\bm{k'}}^{\dag} + v_{\bm{k'}}\hat{\alpha}_{\bm{-k'}}) \ket{\varnothing_{\alpha}}.	
\end{align}
Working out (\ref{eq:rh_dens}) one obtains two kinds of terms, proportional to vacuum state and proportional to 2-magnon states,
\begin{align}
&\ket{\phi \left(\bm{r},t\right) } =
\frac{1}{N}\sum_{\bm{k}} u_{\bm{k}}^2 e^{-i(\bm{k}\bm{r} + \omega_{\bm{k}}t/\hbar)} \ket{\varnothing_{\alpha}} + \nonumber
\\
&+ \frac{1}{N}\sum_{\bm{k},\bm{k'}} u_{\bm{k'}}v_{\bm{k}} e^{-i(\bm{k}\bm{r}_i - \bm{k'}\bm{r}_0 + \omega_{\bm{k'}}t/\hbar)} \sqrt{1 + \delta_{\bm{k'},\bm{-k}}}\ket{\alpha_{\bm{k'}}, \alpha_{\bm{-k}}}.	
\end{align}
Combining the above result with its Hermitian conjugate as in right-hand side of (\ref{eq:rho_def}), and using parity of dispersion relation, we arrive at desired density profile,
\begin{align}
\rho\left(\bm{r},t\right) = &\frac{1}{N^2} \left( \left\vert \sum_{\bm{k}} u_{\bm{k}}^2  e^{i (\bm{k}\bm{r} - \omega_{\bm{k}}t/\hbar)}  \right\vert^2 \right. \nonumber \\
&+ \left. \left\vert \sum_{\bm{k}} u_{\bm{k}} v_{\bm{k}}  e^{i (\bm{k}\bm{r} - \omega_{\bm{k}}t/\hbar)}  \right\vert^2 + \sum_{\bm{k},\bm{q}} u_{\bm{k}}^2 v_{\bm{q}}^2 \right).
\end{align}
The last term, $\sum_{\bm{k},\bm{q}} u_{\bm{k}}^2 v_{\bm{q}}^2$, is just a constant number describing level of quantum fluctuations present in the system. Since we are interested in the dynamics, in the figures we show only those parts of the density profile that are time-dependent,
\begin{equation}\label{eq:rho}
{\rho}\left(\bm{r},t\right) \sim \left\vert \sum_{\bm{k}} u_{\bm{k}}^2  e^{i (\bm{k}\bm{r} - \omega_{\bm{k}}t/\hbar)}  \right\vert^2 + \left\vert \sum_{\bm{k}} u_{\bm{k}} v_{\bm{k}}  e^{i (\bm{k}\bm{r} - \omega_{\bm{k}}t/\hbar)}  \right\vert^2.
\end{equation}
Thus, we observe that the obtained density profile corresponds to that of a superposition
of the plane waves (i.e. it is a wave packet), whose respective weight depends on the dispersion relation $\omega_{\bm{k}}$.

Having obtained the magnon density profile, it is interesting to compare it against the spin-spin correlation function, or spin dynamical structure factor, which is typically used and measured in the condensed-matter community.
A standard definition of the spin dynamical structure in the time domain and real space is
\begin{subequations}
\begin{align}
S^{\alpha\beta}\left(\bm{r},t\right) &= \bra{\varnothing} \hat{S}^{\alpha}_{\bm{r}_i} (t) \hat{S}^{\beta}_{\bm{r}_0}\ket{\varnothing},
\\
 \hat{S}^\alpha_{\bm{r}_i} \left(t\right) & = e^{i\frac{\hat{H}}{\hbar}t}  \hat{S}^\alpha_{\bm{r}_i}  e^{-i\frac{\hat{H}}{\hbar}t},
\end{align}
\end{subequations}
where $\bm{r} = \bm{r}_i - \bm{r}_0$ (see e.g. Ref.~\onlinecite{Auerbach1994}).
Performing the similar, though even simpler, analytical transformation as above we obtain
that in the bosonic language, and in the LSW approximation, the transversal components~\footnote{The longitudinal component corresponds to the two-magnon signal, cf.~\onlinecite{Fuhrman2012} and is not of interest here} of the time-domain and real-space spin dynamical structure factor read
\begin{align}
S^{xx}\left(\bm{r},t\right) &= \frac{1}{4N} \sum_{\bm{k}} (u_{\bm{k}} + v_{\bm{k}})^2 e^{i (\bm{k}\bm{r} - \omega_{\bm{k}}t/\hbar)}, \\
S^{yy}\left(\bm{r},t\right) &= \frac{1}{4N} \sum_{\bm{k}} (u_{\bm{k}} - v_{\bm{k}})^2 e^{i (\bm{k}\bm{r} - \omega_{\bm{k}}t/\hbar)},
\end{align}
where we recognize the typical form of the Bogolyubov factors~\cite{Fuhrman2012}.
Thus, we observe that the magnon density profile discussed in this paper is closely related
to the usual form of the spin dynamical structure: the absolute values of the transversal components
of $S^{\alpha \beta}\left(\bm{r},t\right)$ provide qualitatively similar information to the magnon density profiles. In particular, the two contrasting patterns between FM and AF cases discussed in this paper are also obviously encoded in the spin dynamical structure factor. However, as the density profiles are readily measured in atomic systems~\cite{2009KarskiScience,2011WeitenbergNature,2013FukuharaNature}, we decided to show these in this paper. This justifies the choice of the correlation function made by Eq.~\eqref{eq:rho_spin}.

\section{Results}
The main result, presented in Fig.~\ref{fig:1}, is the distinct density profile of spin excitations created in the FM and AF background. While the FM case resembles a quantum walker with a square pattern that reflects the underlying lattice structure, the AF case resembles a classical wave with a circular pattern that is quasi-ignorant of the underlying microscopic lattice. In the momentum space picture, this can be understood by considering the respective magnon dispersion relations. For the FM [Fig.~\ref{fig:1}(a)] the dispersion is quadratic ($\propto$ $\bm{k}^2$) near $\Gamma$, and its largest slope, and therefore the highest magnon velocity, stems from other parts of the Brillouin zone. Since the local spin flip is composed of all momenta in the Brillouin zone, its spread velocity is dominated by those fast components, which reflect the lattice structure. By striking contrast, for the AF case the dispersion is linear ($\propto$ $|\bm{k}|$) near $\Gamma$, where it also has its largest slope. Therefore the spread of the spin-flip excitation is dominated by the momentum-space region near $\Gamma$, with its emergent isotropic symmetry at long wavelengths, ignorant of the underlying square lattice On top of that, the quantum fluctuations encoded in the Bogolyubov transformation additionally put a stronger focus on the $\Gamma$ point region for the AF case, since the coherence factors modulate the contributions from different momentum-space regions to the wave packet dynamics (for details see Appendix~\ref{appx:density-t0}). As an important consequence of these arguments, the observed striking differences in the spacetime dynamics between FM and AF are expected to be largely insensitive to the details of the prepared initial state, as long as it is sufficiently localized, and also insensitive to the details of the Hamiltonian realizations.

We now elucidate the emergence in real time of the patterns discussed here. In Fig.~\ref{fig:2} we show snapshots of the time evolution of the density profile for the FM (top row) and AF (bottom row) cases. Interestingly, in both cases the characteristic density profiles emerge quickly between the earliest times ($2.0 \hbar/J$) and the next snapshots shown here at $22.5 \hbar/J$. At increasingly longer times, we find the development of self-similar patterns for both FM and AF cases, with a speed of expansion that remains constant over time. This observation is in line with the above momentum-space interpretation: the wave fronts of the density profiles evolve according to the fastest available velocities in the respective wave packets. 

\begin{figure}
	\centering
	\includegraphics[width=\columnwidth]{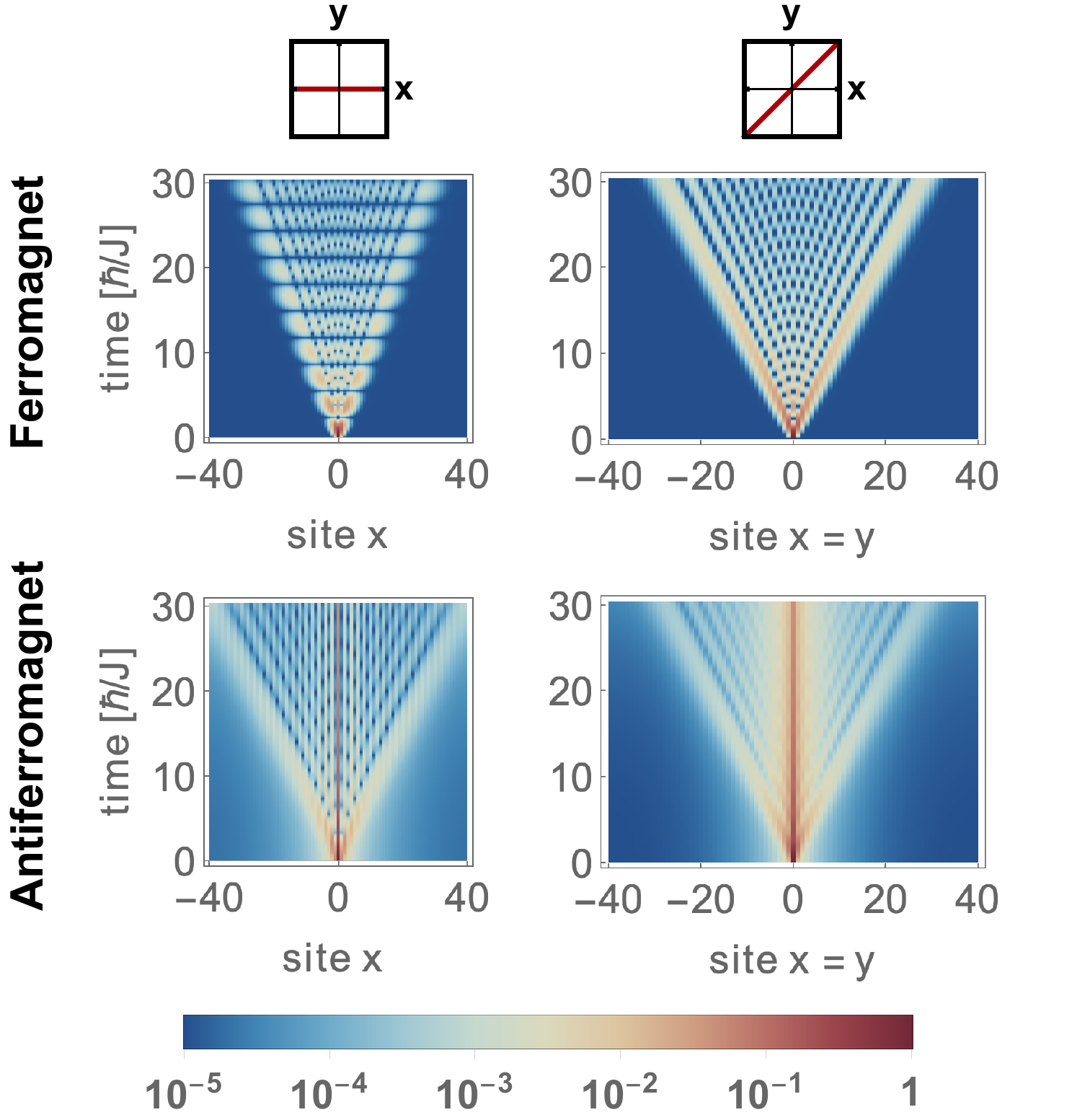}    
	\caption{{\bf Light-cone-like structures along selected real-space cuts.} Spatial cuts of density profiles along $x$ (left panels) and along $x-y$ diagonal (right panels) for the FM (top) and AF (bottom) cases, respectively.
	}\label{fig:3}
\end{figure}

In order to highlight this constant-velocity spreading and to also investigate some more subtle differences between the FM and AF, we present in Fig.~\ref{fig:3} the density profiles along selected real-space cuts in the 2D lattice as functions of time. In the FM case, for the spreading along the $x$ direction, and by symmetry also along the $y$ direction (not shown), one observes a well-known light-cone-like structure \cite{lieb_finite_1972}. In the diagonal direction, one also observes a similar light cone but with a velocity that is larger by a factor of $\sqrt{2}$, again highlighting the momentum-space picture discussed above, and leading to the characteristic square-shaped density profile of the quantum walker. On top of these overall features, we also note that the highest density is found at the edge of the light cone for the FM case. This latter more detailed feature is in stark contrast to the AF case. In the AF, the highest density remains at the center of the excitation. Moreover, as already discussed, the classical-wave-like circular-shaped picture emerges because the light cone spreads isotropically, i.e., equally fast both along the crystal axes and along the diagonal.

\section{Discussion}
\begin{figure}
	\centering
	\includegraphics[width=\columnwidth]{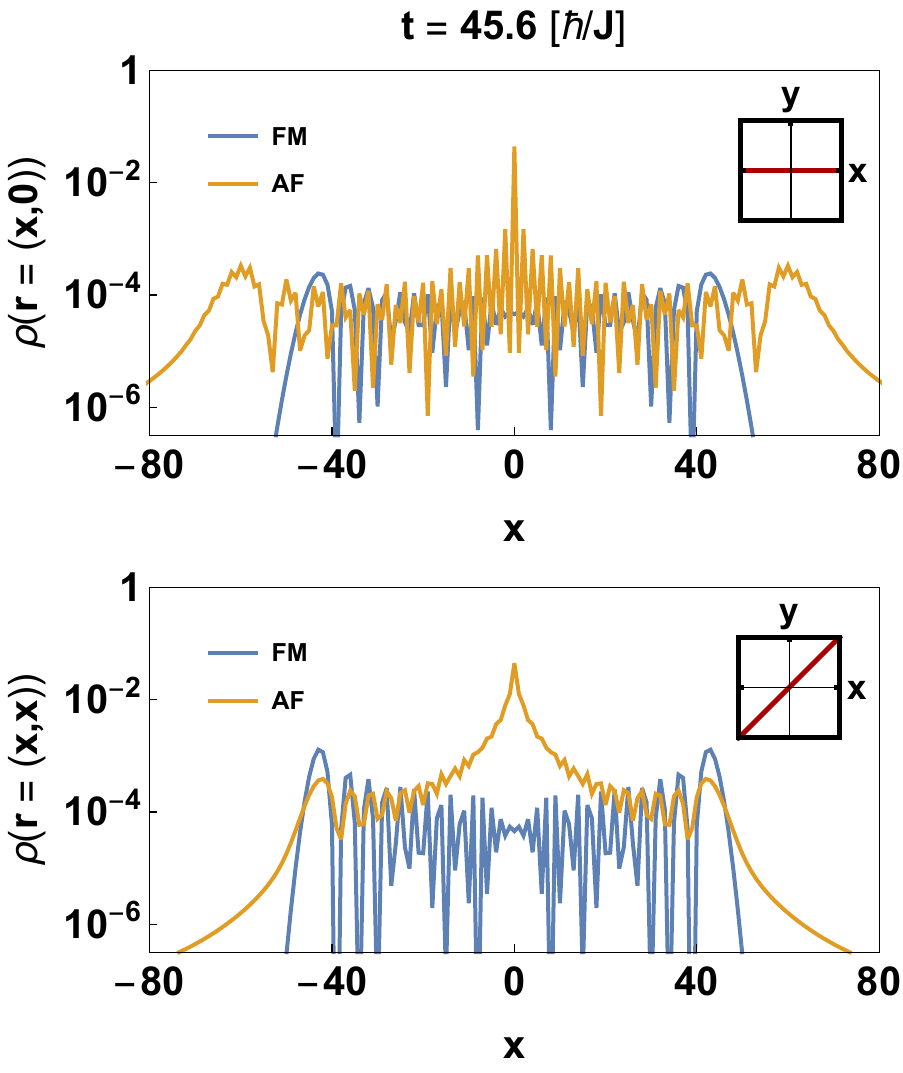}    
	\caption{{\bf Magnon density profile ${\rho}\left(\bm{r},t\right)$ at time $t=45.6 [\hbar / J]$.} Top (bottom) panel shows cuts along the $OX$ (diagonal) directions for the FM (blue) and AF (yellow) ground state, respectively.
	}\label{fig:6}
\end{figure}

Let us firstly comment on why the spatial magnon density profiles resemble the observed light-cone-like structures. A priori, this may seem unexpected since the magnon propagation via the evolution operator $e^{-i \hat{H} t}$ should cause the magnon wave function to be non-zero on all lattice sites for all times $t>0$.
This apparent paradox is resolved by the fact that the light cones are not sharply defined---a magnification of the magnon density profile shows that the probabilities of detecting a magnon excitation outside the cone is indeed non-zero, even though it quickly decays (Fig.~\ref{fig:6}).
This is true not only in the AF but also in the FM case, where this is caused by the fact that at a particular time $t$ the higher order terms in the expansion of the magnon evolution operator $e^{-i \hat{H} t} \approx 1 + (-i \hat{H} t) + \frac12 (-i \hat{H} t)^2 + \ldots$ are never completely suppressed.

\begin{figure}
	\centering
\includegraphics[width=1.0\columnwidth]{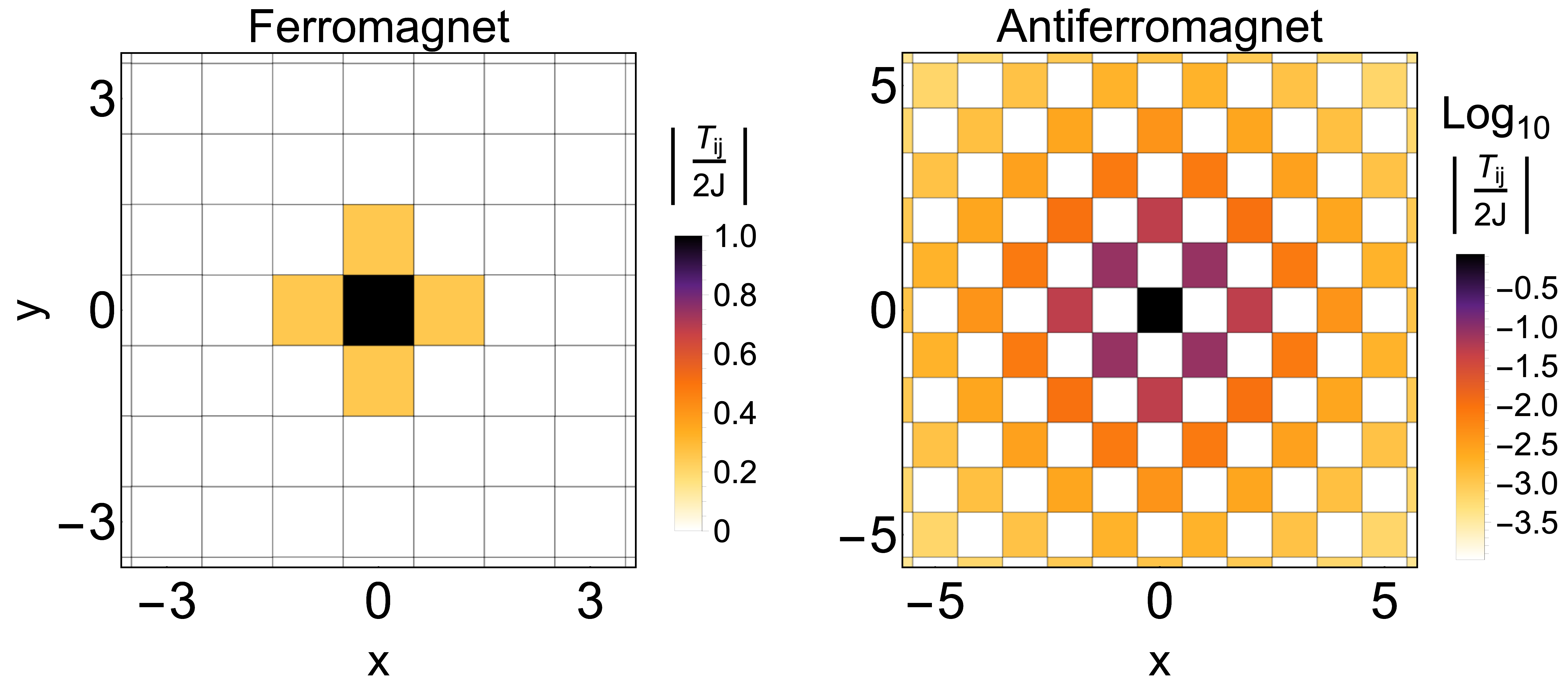}
	\caption{{\bf Real-space effective hopping interpretation of quantum walk versus classical wave.}
 The tunneling amplitudes $\left\vert T_{ij}/2J \right\vert$ shown on a square lattice for FM (left panel) and AF (right panel) groundstate, respectively.}\label{fig:4}
\end{figure}

Next, let us provide a comprehensive real-space understanding of the observed cone-like spreading and real-space structures. To this end we will split the problem into three steps. 

(i) Explanation of a finite density of magnons along the abscissa axis already at time $t=0$ in the AF case (bottom panel in Fig.~\ref{fig:3} and Fig.~\ref{fig:a1-rho}). This can be clarified by investigating consequences of a single spin flip (a magnon {\it before} Bogolyubov transformation) at the initial time. It is clear that the single Bogolyubov magnon at site ${\bf r}_0$, which is a Fourier-transformed eigenstate of the system, is a superposition of spin flips whose real-space distribution decays with distance from site ${\bf r}_0$ (for more details see Appendix). This originates in the fact that the Hamiltonian makes it energetically favorable to cluster the magnons, which are always present even in the ground state of the AF, near the additionally created spin flip at time $t=0$. 

(ii) Unraveling why the spatiotemporal propagation of a magnon in the FM (AF) resembles the
quantum walk (classical-wave) case, respectively. This is achieved by investigating the hopping amplitudes of a single magnon in real space. In the FM case, the situation is clear, since the Hamiltonian can be easily written in terms of bosons hopping on a lattice, $\hat{H}= \sum_{i j} T_{i j} \hat{a}_i^{\dag} \hat{a}_j$ with hopping amplitudes $T_{i j} = 2J \left( \delta_{\bm{r}_i - \bm{r}_j} + \frac{1}{4}\delta_{\bm{r}_i - \bm{r}_j \pm \hat{\bm{x}}} + \frac{1}{4}\delta_{\bm{r}_i - \bm{r_j \pm \hat{\bm{y}}}} \right)$, being nonzero only for nearest-neighbor sites and within a given site [Fig.~\ref{fig:4}(a)]. This effective hopping matrix structure exactly defines a quantum walk problem. On the other hand, the AF Hamiltonian needs to be rewritten more carefully since the Bogloyubov transformation is required to eliminate magnon pair creation or annihilation terms. Consequently, the real-space representation is achieved by Fourier transforming the AF Hamiltonian written in the Bogolyubov magnons. We obtain $\hat{H}= \sum_{i j} T_{i j} \hat{\alpha}^\dag_i \hat{\alpha}_j $, with $T_{i j} = \frac{1}{\sqrt{N}}\sum_{\bm{k}} \omega_{\bm{k}} e^{i \bm{k}(\bm{r}_i-\bm{r}_j)}$. From this it is clear that for the AF case, the tunneling amplitudes $T_{i j}$ are nonzero to all sites on the same sub-lattice, though they decay with distance as $|\bm{r}_i-\bm{r}_j|$ [Fig.~\ref{fig:4}(b)]. Thus, one observes emergence of an isotropic circular shape of the spatiotemporal magnon density profile, just as in the case of a classical wave.

Interestingly, the origin of such a particular long-range hopping amplitude in the AF lies in the interplay between the Bogolyubov transformation and the nearest-neighbor pair creation and annihilation present in the Hamiltonian written in the language of the original Holstein-Primakoff magnons, {\it i.e.}, before Bogolyubov transformation. The crucial observation here is that when a bosonic particle residing on site $j$ is Bogolyubov-transformed to a bosonic hole, then the latter can reside on any site of the other sub-lattice than site $j$, but with a decaying probability with distance from site $j$. This is because we have the relation $\hat{a}_{{\bf r}_j}^\dag \rightarrow \sum_{\bf l} G({\bf r}_{lj}) \hat{\alpha}_{{\bf r}_l} $, where $G({\bf r}_{lj}) = \sum_{\bf q} \exp [i {\bf q} ({\bf r}_l - {\bf r}_j)] v_{-{\bf q}}$ and $v_{\bf q}$ is the relevant Bogolyubov coefficient. Thus, when the nearest-neighbor magnon pair creation is Bogolyubov-transformed to the creation of a hole and a particle, it yields nonzero transition coefficients $T_{i j}$ connecting all sites on the same sub-lattice but with decaying values with increasing distance $|\bm{r}_i-\bm{r}_j|$, as discussed above. 

(iii) Unfolding a relatively large, steady-in-time probability for detecting the excitation at the initial position in the AF case, as clearly visible by comparing the FM and the AF light cone distributions in Fig.~\ref{fig:3}. This specific dissimilarity is a direct consequence of the differences between the creation of magnons at the initial time in the FM and AF cases, already discussed above. In the FM state at time $t=0$ there is just one point in space where the magnon is created and thus the magnon wave function spreads relatively fast from site ${\bf r}_0$. In the AF case, on the other hand, the magnon is initially created at several sites, and hence the probability of finding the magnon at site ${\bf r}_0$ decreases relatively slowly with time.

Before concluding, a few words are in order to discuss the validity of the LSW approximation in the context of a single-spin flip excitation in the Heisenberg model. 
First of all, for the FM case, the LSW is exact for the 0- and 1-magnon sectors that are relevant here
\cite{Auerbach1994}.
For the AF case, on the other hand, the crucial feature of the magnon dispersion is its linearity at small wavevectors, leading to the classical-wave pattern in the magnon density profile. This linearity is preserved even under magnon-magnon 
interactions~\cite{Manousakis1991}.
Furthermore, the magnon excitation becomes long-lived in this long-wavelength 
limit~\cite{Harris1971, Manousakis1991}
which means that the dominant wave-front features are correctly captured by LSW. 
We also note that the single-magnon excitation sector probed in our setup is profoundly different from the two-magnon excitation sector with regard to the role of magnon-magnon scattering.
The two-magnon sector is probed for instance in two-magnon Raman scattering \cite{canali_theory_1992} or directed spin transport under external fields \cite{sentef_spin_2007}.
Finally, to further support the validity of the LSW approximation, we have compared the LSW results to the exact numerical dynamics of the Heisenberg Hamiltonian on a small lattice.
The results show full agreement for the FM case and corroborate our physical conclusions for the AF case.
Thus, the results obtained here using the LSW approach remain valid beyond that approximation.
For further details see Appendix~\ref{appx:ed}.

\section{Conclusion}
In conclusion, we have presented an intriguingly simple way of characterizing prototypical magnetic ground states in quantum materials by their spacetime dynamics. We have shown that the ferromagnetic quantum walker is intimately tied to the quadratic magnon dispersion, whereas the antiferromagnetic walker has an emergent classical dynamics, tied to its linear magnon dispersion like for classical acoustic sound or water waves. These deep connections, while not being too surprising after all, open important possibilities for studying the important quantum-magnetic properties of materials, besides the obvious potential realizations in quantum simulators. In particular, the subtle magnetic ground states in recently discovered two-dimensional van der Waals materials with CrI$_3$ \cite{huang_layer-dependent_2017} as a truly atomically thin ferromagnet, would make for interesting test objects of our predictions, provided that real-space and real-time imaging techniques can be pushed accordingly. Similarly, there are some well-known realizations of quasi two-dimensional Heisenberg antiferromagnets \cite{dalla_piazza_fractional_2015}, and light-cone spreading has only recently been simulated in such systems \cite{fabiani_ultrafast_2019}. A potential experimental probe is time-resolved resonant inelastic X-ray scattering, as proposed for instance in \cite{wang_real-space_2014} and demonstrated in \cite{dean_ultrafast_2016}. A further intriguing avenue for spacetime imaging is the opportunity to monitor Floquet-engineered magnetic exchange interactions \cite{mentink_ultrafast_2015,kennes_floquet_2018}, which in turn would affect the light-cone-like dynamics \cite{kalthoff_floquet-engineered_2019}. We finally mention the intriguing possibility to investigate anomalous spin diffusion, similar to the anomalous charge diffusion reported in Ref.~\onlinecite{mitrano_anomalous_2018}), through  spacetime dynamics.

\section*{Acknowledgments}
Financial  support  by  the  DFG through  the  Emmy  Noether  program  (SE  2558/2-1)  is gratefully  acknowledged.   We  kindly  acknowledge  support  by  the  (Polish)  National  Science  Centre  (NCN, Poland)  under  Projects  No.2016/22/E/ST3/00560 (PW and KW),   2016/23/B/ST3/00839 (KW),   and 2016/22/E/ST2/00555 (TS).

\appendix
\section{Magnon density profiles at initial time}
\label{appx:density-t0}

\begin{figure}[htbp]
	\centering
	\includegraphics[width=\columnwidth]{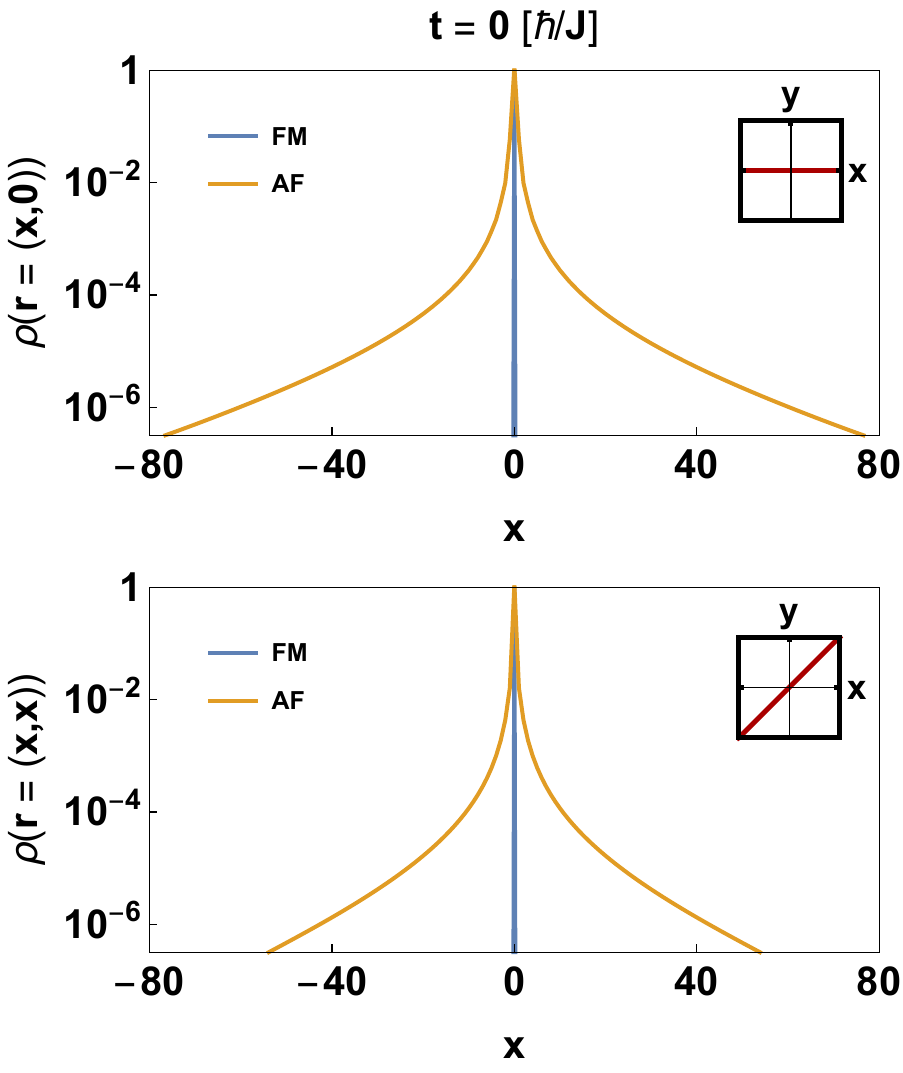}    
	\caption{{\bf Magnon density profile ${\rho}\left(\bm{r},t\right)$ at time $t=0$.} Top (bottom) panel shows cuts along the $OX$ (diagonal) directions for the FM (blue) and AF (yellow)ground state, respectively.}
	\label{fig:a1-rho}
\end{figure}

\begin{figure}[htbp]
	\centering
	\includegraphics[width=\columnwidth]{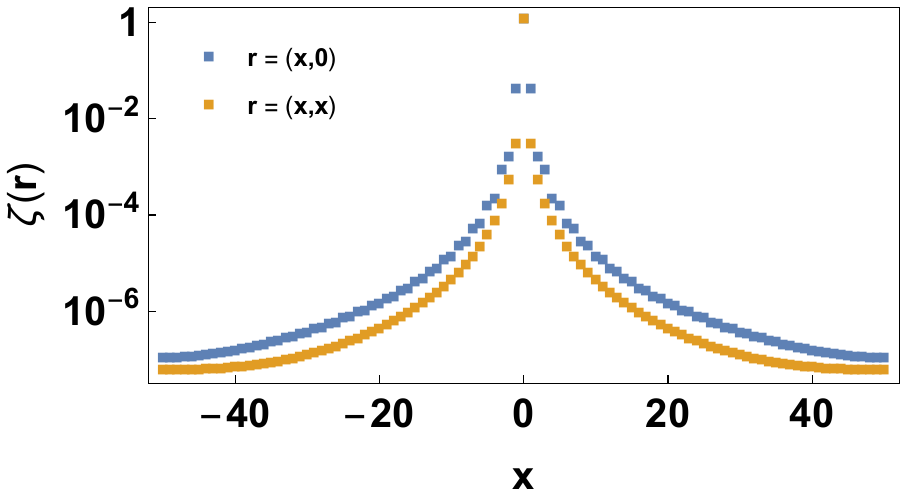}    
	\caption{{\bf Magnon density profile $\zeta\left(\bm{r}\right)$ at time $t=0$ of a single Bogolyubov boson in the AF ground state.}
    Blue and yellow dots show results along the $OX$ and diagonal direction, respectively.
	}
	\label{fig:a1-zeta}
\end{figure}

The magnon density profile ${\rho}\left(\bm{r},t\right)$ at time $t=0$, i.e. at the time that a single spin flip is created, is shown in Fig.~\ref{fig:a1-rho}. As discussed in the main text, we observe that in the antiferromagnetic case a creation of a single spin flip at site ${\bf r}_0$ leads to a whole cloud of magnons being instantaneously created around this site. By contrast, this is not the case for a ferromagnet, for which the creation of a single spin flip at site ${\bf r}_0$ corresponds to just one single magnon instantaneously created at the same site and no magnons on other sites. We explain this phenonemon in two steps. (i) We express the creation operator of a single magnon at site ${\bf r}_0$ in terms of Bogolyubov magnons $\hat{\alpha}_{\bf r}$. It then turns out that, creating a single magnon at site ${\bf r}_0$ is equivalent to the creation of a cloud of Bogolyubov magnons centered around ${\bf r}_0$ and with an exponentially decaying probability of finding them away from ${\bf r}_0$. This is due to the combination of the Bogolyubov transformation and the relation 
\begin{align}
|a_{{\bf r}_j} \rangle = \sum_{l} \sum_{\bf q} u_{\bf q}\exp [i {\bf q} ({\bf r}_l - {\bf r}_j)]  |\alpha_{{\bf r}_l} \rangle 
\label{eq:111}
\end{align}
where $u_{\bf q}$ is the coefficient of the Bogolyubov transformation (\ref{eq:bog}). 
(ii) It occurs that the density of magnons in a single Bogolyubov particle $|\alpha_{\bf  r}\rangle$ also decays exponentially when going away from site ${\bf r}_0$. Indeed the magnon density profile $\zeta\left(\bm{r}\right)$ of a single Bogolyubov boson
at time $t=0$ defined as
\begin{align}
\zeta\left(\bm{r}\right) = &\bra{\varnothing_{\alpha}} \hat{\alpha}_{\bm{r}_0}  \hat{a}_{\bm{r}_i}^{\dag} \hat{a}_{\bm{r}_i} \hat{\alpha}_{\bm{r}_0}^{\dag} \ket{\varnothing_{\alpha}} - \bra{\varnothing_{\alpha}} \hat{a}_{\bm{r}_i}^{\dag} \hat{a}_{\bm{r}_i} \ket{\varnothing_{\alpha}} \nonumber \\ &= \left| \frac{1}{N} \sum_{\bm{k}} e^{i\bm{kr}} u_{\bm{k}} \right|^2 + \left| \frac{1}{N} \sum_{\bm{k}} e^{i\bm{kr}} v_{\bm{k}} \right|^2
\label{eq:222}
\end{align}
is displayed in Fig.~\ref{fig:a1-zeta}.

Altogether, the relation between the single spin flip created at site ${\bf r}_0$ of the AF ground state and the resulting distribution of magnons in such an excited state is a function of the product of the above two equations.
This leads to the calculated magnon density profile (\ref{eq:rho}) at time $t=0$ and to the observed magnon density profile presented in Fig.~\ref{fig:a1-rho}. The intuitive understanding of this result is as follows: the AF Hamiltonian makes it energetically favorable to cluster the magnons (which are already present in the AF ground state) near the additionally created spin flip at time $t=0$, as already stated in the main text of the paper.

\section{Comparison between linear spin-wave theory and exact dynamics}
\label{appx:ed}

\begin{figure*}[htbp]
	\centering
	\includegraphics[width=0.9\textwidth]{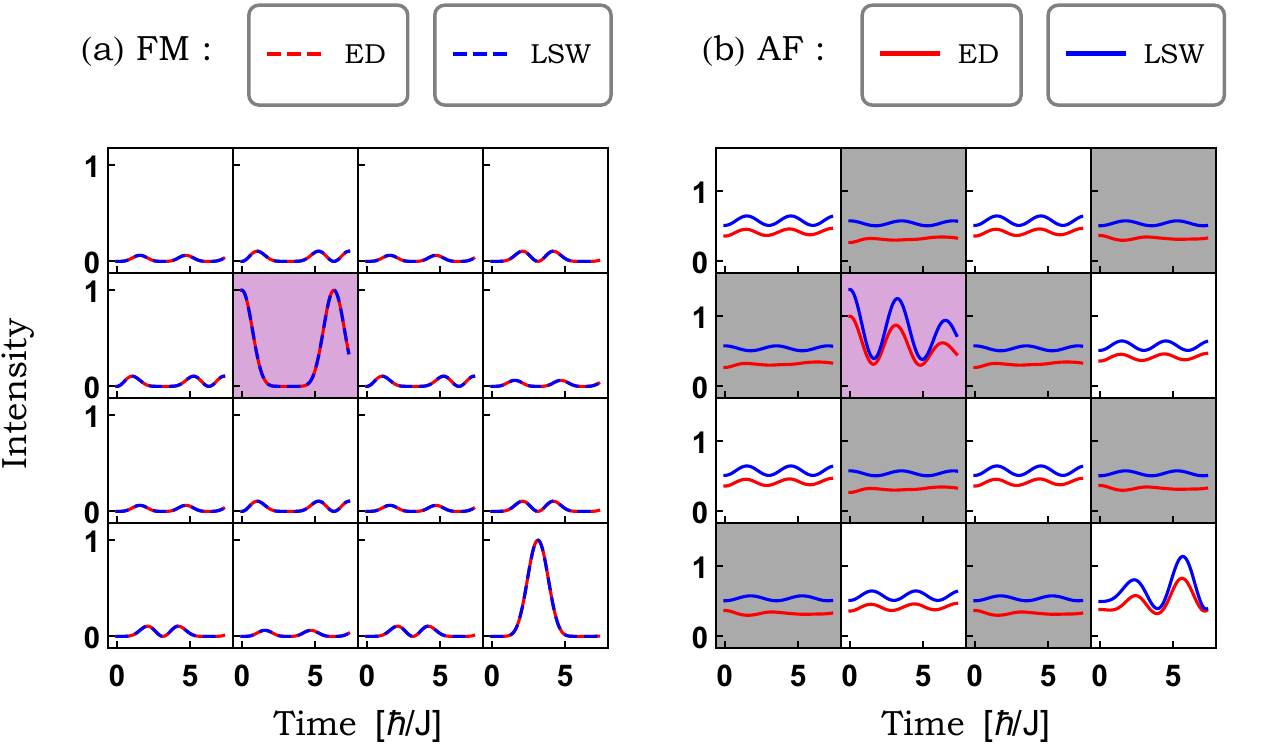}    
	\caption{{\bf Propagation of a single spin flip
	created in the ground state of the
	Heisenberg model (ED) compared with magnon density profile $\rho\left(\bm{r}, t\right)$ of a single Holstein-Primakoff boson in the vacuum state of Bogolyubov bosons (LSW).} 
	Both results
	are shown for a  $4\times 4$ lattice with periodic boundary conditions. At the initial time $t = 0$ the spin-flip excitation is created on the site in the second row and second column of the grid (highlighted by colored background). (a) Ferromagnetic case. (b) Antiferromagnetic case. 
	Here, the panels corresponding to sites on the other antiferromagnetic sublattice with respect to the initial excitation have a grey-shaded background.
	} \label{fig:8}
\end{figure*}

We briefly comment on the accuracy of employing linear spin-wave theory (LSW) for the dynamics after a single spin-flip. To this end, we perform benchmark calculations for a small 4 $\times$ 4 lattice with periodic boundary conditions, comparing LSW with exact diagonalization (ED). The ED results have been obtained by numerically solving the Schr\"odinger equation for the full quantum state using QuTiP 4.5.0~\cite{johansson_qutip_2012}.
We have further used functionality from NetKet 2.1b1~\cite{carleo_netket_2019} to setup the system and initial states.

Figure~\ref{fig:8} shows the comparison between LSW and ED.
In both cases, the initial state is prepared by applying a single-site spin-flip, i.e., the $\hat S^-_{l}$ operator for fixed site $l$, to the respective ground state.
For the FM case [Fig.~\ref{fig:8}(a)], the LSW and ED results are exactly identical, proving that LSW is exact both for the ground state and for a single spin flip 
excited state
in the FM Heisenberg model. The underlying reasons are (i) the absence of quantum fluctuations in the ground state (the fully polarized classical ground state is the exact vacuum), and (ii) the fact that magnon-magnon scattering in the FM only occurs for two magnons scattering into two other magnons. Therefore a single magnon does not find any scattering partner, and single-magnon excitations can propagate ballistically in the FM. 

For the AF case [Fig.~\ref{fig:8}(b)] we do find some deviations, which is expected. First of all, we note that the effective exchange coupling within LSW needs to be corrected here using the so-called Oguchi correction factor\cite{oguchi_theory_1960}, $J_{\text{eff}} \approx 1.158 J$, which is a well-known quantum-fluctuation correction stemming from normal ordering of quartic terms in the spin-wave Hamiltonian. Once this is taken into account, the results on the 
antiferromagnetic
sublattice on which the excitation is created do agree qualitatively between LSW and ED.
The results on the other sublattice are out of phase,
which to our understanding is due to the fact that in the ED calculations the ground state does not have a broken symmetry and hence the ED calculations do not differentiate between
the two antiferromagnetic sublattices. Since the ground state of the 2D Heisenberg model in the thermodynamic limit is widely believed to have a broken symmetry~\cite{Reger1988, Chakravarty1989}, we suggest that the LSW
may actually better reflect the exact case of an infinite lattice than the ED performed on a small cluster. 

We further note that the magnon occupation is larger than unity initially on the site where the spin flip occurs.
This is due to the fact that the bosonic occupation on this site is not restricted to unity within our calculations.
Such a constraint is only fulfilled for the number of magnons averaged over the entire lattice in the LSW calculations.
Thus, the creation of a boson at $t=0$ happens on top of a background that already has a partial bosonic occupation locally, leading to the magnon density becoming larger than unity. Importantly though, 
this relatively small
quantitative discrepancy between ED and LSW does not invalidate the key result of the main text. 

\bibliographystyle{apsrev4-1}
\bibliography{bibliography}


\end{document}